\documentclass{iaus}
\usepackage{graphicx}

\title[Abundance Inhomogeneities in Star Clusters] 
{Origin of Star-to-Star Abundance Inhomogeneities in Star Clusters}

\author[Jan Palou\v s, Richard W\" unsch, Guillermo TenorioTagle \&
  Sergyi Silich]   
{Jan Palou\v s$^1$,
Richard W\" unsch$^2$, \\ 
Guillermo Tenorio-Tagle$^3$, 
\and Sergyi Silich$^3$
}

\affiliation{$^1$Astronomical Institute, Academy of Sciences of the
  Czech Republic, \\ Bo\v cn\' \i \ 1401, 140 31 Prague 4, Czech
  Republic
\\ email: {\tt palous@ig.cas.cz} \\[\affilskip]
$^2$Cardiff University, Queens Buildings, The Parade, Cardiff, \\ 
CF24 3AA, United Kingdom\\email: {\tt richard@wunsch.cz}\\
[\affilskip]
$^3$Instituto Nacional de Astrof\' \i sica Optica y
  Electronica, \\ AP 51, 72000 Puebla, M\' exico\\email: {\tt
    gtt@inaoep.mx, silich@inaoep.mx}
}

\pubyear{2008}
\volume{254}  
\pagerange{119--126}
\jname{The Galaxy Disk in Cosmological Context}
\editors{J. Andersen, J. Bland-Hawthorn \& B. Nordst\" om, eds.}
\begin{document}

\maketitle

\begin{abstract}
The mass reinserted by young stars of an emerging massive compact
cluster shows a bimodal hydrodynamic behaviour. In the inner part of
the cluster, it is thermally unstable, while in its outer parts it
forms an out-blowing wind. The chemical homogeneity/inhomogeneity of
low/high mass clusters demonstrates the relevance of this solution to
the presence of single/multiple stellar populations. We show the
consequences that the thermal instability of the reinserted mass has
to the galactic super-winds.
\keywords{Galaxy: abundances, Galaxy: Globular clusters: general,
  Galaxy: open star clusters and associations: general, 
galaxies: star clusters, galaxies: starburst}
\end{abstract}

\firstsection 
\section{Introduction}

The open star clusters and stellar moving groups have internally homogeneous
chemical composition. Clusters like the Hyades, Collinder 261, the Herculis
stream or the moving group HR 1614 are chemically unique, distinguishable 
one from the other, showing no pollution from secondary star formation
\cite[(De Silva et al., 2008)]{Silva3}.
The chemical homogeneity of open star clusters like the Hyades 
\cite[(De Silva et al., 2006)]{Silva1}
and Collinder 261 \cite[(De Silva et al., 2007)]{Silva2} proves that they have
been formed out of a well-mixed cloud and that any self-enrichment of
stars did not take place there.

Young and massive stellar clusters, frequently called super star
 clusters, are preferentially observed in interacting galaxies. Their
 stellar mass amounts to several million M$_{\odot }$ within a region
 less than a few parsecs in diameter. They represent the dominant mode
of star formation in starburst galaxies. Their high stellar
 densities resemble those of globular clusters, where several
 stellar populations have been observed \cite[(Piotto, 2008)]{GP}.

To explain the presence of multiple stellar generations in 
globular clusters, the slow wind emerging from a first generation of
fast rotating massive stars is invoked by 
\cite[Decressin et al. (2007)]{TDetal}. (See also the review by 
\cite[Meynet (2008)]{Meynet} in this volume.) The authors argue that
the fast rotating massive stars function as a filter separating the 
H-burning products from later products of He-burning. However, it is not
clear why all the massive stars rotate fast, or why the slow wind 
produced by stellar rotation is just retained inside the potential well 
of the stellar cluster.

An alternative  solution, how to form the  second generation of  stars  
in massive star clusters, is proposed in models of star cluster winds
described by 
\cite[Tenorio-Tagle et al. (2007)]{GTTetal}, 
\cite[W\" unsch et al. (2007)]{RWetal1},
and \cite[W\" unsch et al. (2008)]{RWetal2}. 
There, we argue that a critical mass of a cluster exists, below which the 
single-mode hydrodynamical solution to the cluster winds applies. 
Such clusters
should have one stellar generation only, and show strong winds corresponding
to the momentum and energy feedback of all their stars. The clusters above the
critical mass should follow the bi-modal solution to their winds, where only 
the outer skin of the cluster participates in the wind. Their inner parts are 
thermally unstable, and hence being the potential places of secondary 
star formation.   

\section{The Model}

\begin{figure}[b]
\begin{center}
 \includegraphics[width=5.1in]{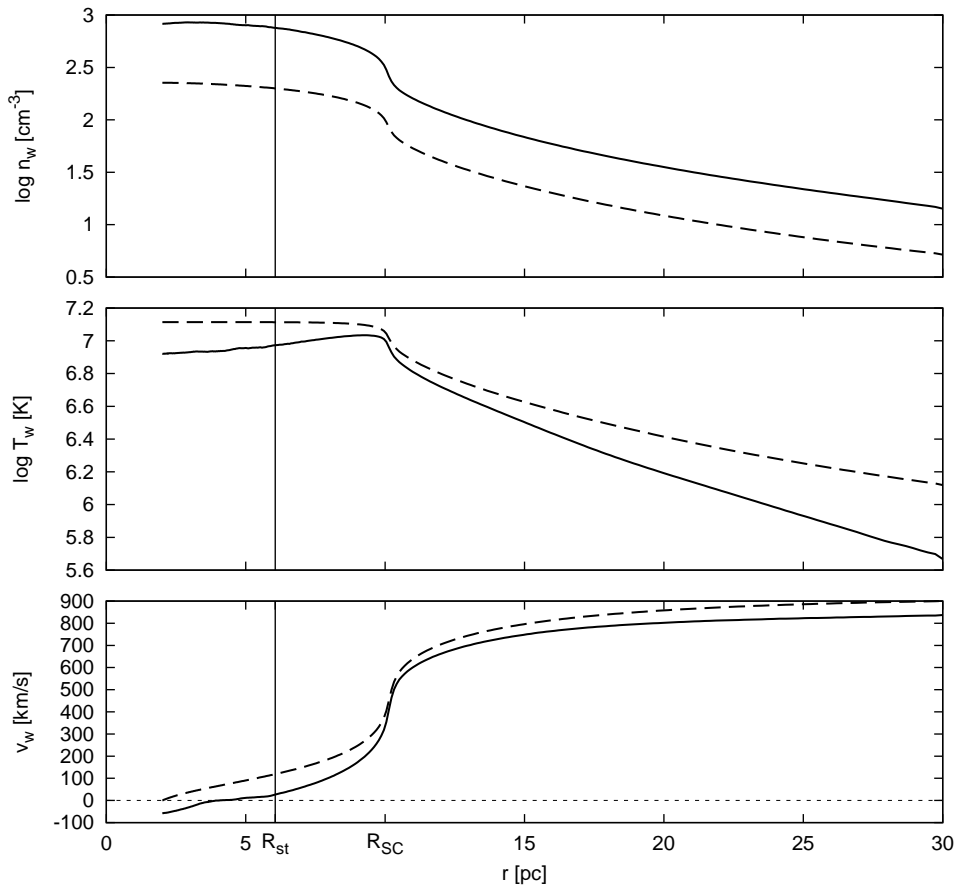} 
 \caption{Radial profiles of particle density, temperature and velocity for 
single-modal (dashed lines) and bi-modal (solid lines) solutions.}
   \label{fig1}
\end{center}
\end{figure}

\begin{figure}[t]
\begin{center}
 \includegraphics[width=5.1in]{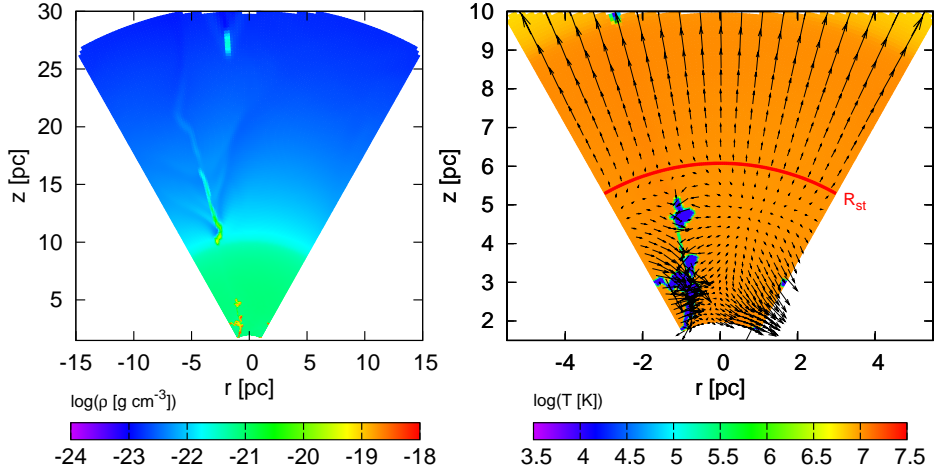} 
 \caption{Bi-modal solution: density distribution (left panel), temperature 
distribution with velocity vectors (right panel) 0.56 Myr after the
beginning of the simulation.}
   \label{fig2}
\end{center}
\end{figure}

The hydrodynamical behaviour of matter reinserted within a star
cluster is described by \cite[Chevalier \& Clegg (1985)]{ChevalClegg}.
In this adiabatic model, the authors assume that all the energy
provided by stellar winds and supernovae is thermalized in random
collisions of the shock waves creating a gas of temperatures $T > 10^7$
K. Chevalier \& Clegg's stable solution shows almost constant density
and temperature inside of the cluster. A mild outward pressure
gradient drives a cluster wind with a radially  increasing velocity
reaching the sound speed $c_{SC}$ at the cluster surface and
approaching $2 c_{SC}$ at infinity. The run of wind particle
density, temperature and wind velocity in this single-mode model is
shown in Fig. 1. With the adiabatic model the 
more massive clusters produce the more powerful wind, taking away all
the elements produced in stars. Thus, any inhomogeneity in stellar
chemical composition reflects the abundance distribution in the
parental star-forming nebula. If the original cloud is well mixed,
according to the adiabatic, single-mode model of  
\cite[Chevalier \& Clegg (1985)]{ChevalClegg},
the star cluster will have one population of the same chemical
composition.
   
The assumption that all the mechanical energy of winds and supernovae
is thermalized need not to be completely true. The efficiency of the
thermalization process $\eta $ depends on the
details of shock-shock collisions and different authors give
different values ranging from 0.01 to 1 \cite[(Melioli, \& Gouveia Dal
  Pino, 2004)]{MeliGou}. The value of $\eta $ may depend on Mach numbers of the
colliding shocks and/or on the chemical composition of the colliding
fronts regulating what part of the mechanical energy is directly
radiated away. The value of $\eta $ may be different in the case of
colliding stellar winds, or colliding supernova explosions. 

The density of the cluster wind $n_w$ depends on the stellar density:
for a cluster of given radius $R_{SC}$, $n_w$ is proportional to the cluster 
mass $M_{SC}$. On the other hand, the radiative cooling rate is
proportional to $n_w^2$ multiplied with a cooling function $\Lambda
(T, Z)$, where $Z$ stands for the chemical composition of the
radiating plasma. Thus, the radiative cooling rate is proportional to
$M_{SC}^2$. However, the cluster mechanical luminosity $L_{SC}$, or
the part that is thermalized $\eta L_{SC}$,
depends linearly on cluster mass $M_{SC}$, which implies that a
critical mass $M_{SC, crit}$ must exist, above which the energy loss
due to radiative cooling and gas expansion exceeds $\eta L_{SC}$. 
Its value depends on $R_{SC}$ and $\eta $. 

Less massive clusters show the
single-mode behaviour. They take away
at least a part of the mechanical luminosity as winds from all the
cluster volume. 
In the adiabatic approximation,
which applies to low mass clusters far below the $M_{SC, crit}$, all
the mechanical energy, which is transformed to heat, is removed as winds. 

For $M_{SC}$ above the critical value, the volume inside of the cluster
is split into two sectors. The inner part of the cluster, inside the
stagnation radius $R_{ST}$, is thermally unstable. There, the
instability leads to fast cooling of small parcels of gas surrounded
by a hot medium. The repressurizing shocks drive the hot gas into
cold gas parcels forming  high density gas concentrations.    
The outer part of the cluster, its skin $R_{ST} < R < R_{SC}$, shows
the out-blowing wind. The run of wind particle
density, temperature and wind velocity in the bi-modal solution of
the star cluster wind is shown in Fig. 1. The single-mode compared to
the bi-modal solution shows wind
of lower density and higher temperature. The velocity of the out-blowing
wind is higher in the case of the single-mode solution compared to
the bi-modal case; the two solutions
have the same velocity at the cluster surface only.  
2D hydrodynamical simulation of cluster winds are described by 
  \cite[W\" unsch et al. (2008)]{RWetal2}. In Fig. 2 we show the
  density and temperature distribution together with velocity vectors 
for a star cluster with $R_{SC}$ = 10 pc, $\eta =1$, and
  $L_{SC}$ = $10^{42}$ erg s$^{-1}$. We can clearly distinguish the
thermally unstable region within the stagnation radius $R_{ST}$ = 6 pc,
and the wind blowing from the outer skin of the cluster. The dense 
concentration, in the thermally unstable part of the cluster, 
shows a small velocity
dispersion relative to the cluster; most of these concentrations 
are unable to leave
the cluster. Thus they accumulate in the inner volume, becoming
potential places of secondary star formation.     

\section{Summary}

We propose a bimodal solution, where in the central 
part of a massive star cluster, a thermally unstable region forms
(see Fig. 2).
In this region, the thermal instability creates cold regions surrounded by 
hot medium imploding into them. The high-velocity wing of 
broad spectral lines
\cite[Gilbert, \& Graham (2007)]{AMG&JRG} observed in SSC may be created
by imploding shock in the vicinity of thermally unstable parcels of gas.

The second generation of stars may be formed out of cold clumps produced 
by thermal instability. 
During the early evolution of a massive cluster, the first Myrs,
the mechanical energy input is dominated by stellar winds
\cite[(Leitherer et al. 1999)]{Leithereretal}. The
efficiency of thermalization $\eta $ may be low in this case, and a
massive cluster may be above $M_{SC, crit}$, since its value is low. 
Later, the importance of winds fades out, and the mechanical input is
dominated by supernovae. This may increase the thermalization
efficiency $\eta $, increasing at the same time the value $M_{SC,
  crit}$. Thus the same
massive cluster, which was initially in the bi-modal situation moves
to single-mode situation. 

The cold parcels of gas form, in the cluster central part 
during the early bi-modal situation, from the winds that 
are enriched by products of H-burning. The later He-burning products
are inserted into the cluster volume when the mechanical energy input
is dominated by supernovae, which may mean that the cluster is in
the single-mode situation, and the wind clears its volume from He burning 
products.  Thus, the thermal instability, which operates during a
few initial Myr in the central part of the cluster, may produce a
second generation of stars enriched by H-burning products. Later, the
cluster moves into the single-mode situation, which means that it is
able to expel the He-burning products.

The feedback of massive stars in super star clusters creates
galactic winds, or super winds, reaching to large distances from 
the parent galaxies, transporting the products of stellar burning into
intergalactic space.  The bimodal solution,
providing a possible explanation of multiple stellar populations in globular
clusters, limits the super winds. During the initial period of 
cluster evolution, when the  stellar winds dominate the mechanical
energy input, the super wind is restricted only to the outer skin of the
cluster, which means that it is rather weak. 
Only later, when supernovae become dominant  in
mechanical energy input, strong super winds blowing out from all
the cluster volume may reach  large distances from their parent galaxies. 
How effective the super winds of super star
clusters can be in transporting the products of stellar evolution into
intergalactic spaces should be discussed in future.

\begin{acknowledgments}
 The authors gratefully acknowledge the support by the Institutional
 Research Plan AV0Z10030501 of the Academy of Sciences of the Czech
 Republic and by the project LC06014 Center for Theoretical
 Astrophysics of the Ministry of Education, Youth and Sports of the
 Czech Republic. RW acknowledges support by the Human Resources and
 Mobility Programme of the European Community under the contract 
 MEIF-CT-2006-039802. This study has been supported by CONACYT-M\'
 exico research grant 60333 and 47534-F and AYA2004-08260-CO3-O1 from
 Spanish Consejo Superior de Investigaciones Cient\' \i ficas.
 The authors express their thanks to Jim Dale for careful reading of
 the text.
\end{acknowledgments}




\begin{thebibliography}{}

\bibitem[Chevalier, \& Clegg (1985)]{ChevalClegg}
    {Chevalier, R. A., \& Clegg, A. W.} 1985,
    \textit{Nature}, 317, 44

\bibitem[De Silva et al.(2006)]{Silva1}
    {De Silva, G. M., Sneden, C., Paulson, D. B., Asplund, M. \&
    Bland-Hawthorn, J.} 2006, 
    \textit{AJ}, 131, 455

\bibitem[De Silva et al.(2007)]{Silva2}
    {De Silva, G. M., Freeman, K. C., Asplund, M., Bland-Hawthorn, J.,
    Bessesl, M. S. \& Collet, R.} 2007, 
    \textit{AJ}, 133, 1161

\bibitem[De Silva et al. (2008)]{Silva3}
    {De Silva, G. M., Freeman, K, Bland-Hawthorn, J., and Asplund,
    M. } 2008,
    in J. Andersen, J. Bland-Hawthorn \& B. Nordst\" om (eds.),
    \textit{The Galaxy Disk in Cosmological Context},
    Proc. IAU Symposium No. 254 (CUP), this volume

    \bibitem[Decressin et al. (2007)]{TDetal}
    {Decresssin, T., Charbonnel, C., \& Meynet, G.} 2007, 
    \textit{A\&A}, 475, 859

\bibitem[Gilbert, \& Graham (2007)]{AMG&JRG}
    {Gilbert, A. M., \& J.~R.~Graham} 2007, 
    \textit{ApJ}, 668, 168

\bibitem[(Leitherer et al., 1999)]{Leithereretal}
    {Leitherer, C., Schaerer, D., Goldader, J. D., Gonz\'
    ales-Delgado, R. M., Robert, C., Foo Kune, D., De Mello, D. F.,
    Devost, D., \& Heckman, T. M.} 1999,
    \textit{ApJS}, 123, 3

\bibitem[(Melioli, \& Gouveia Dal Pino, 2004]{MeliGou}
    {Melioli, C., \& de Gouveia Dal Pino, E. M.} 2004,
    \textit{A\&A}, 424, 817

\bibitem[Meynet (2008)]{Meynet}
   {Meynet, G.} 2008,
    in J. Andersen, J. Bland-Hawthorn \& B. Nordst\" om (eds.),
    \textit{The Galaxy Disk in Cosmological Context},
    Proc. IAU Symposium No. 254 (CUP), this volume

\bibitem[(Piotto, 2008)]{GP}
    {Piotto, G.} 2008,
   \textit{MemSAI}, 79, 3  

\bibitem[Tenorio-Tagle et al. (2007)]{GTTetal}
    {Tenorio-Tagle, G., W\" unsch, R., Silich, S., \& Palou\v s, J.} 2007, 
    \textit{ApJ}, 658, 1196

\bibitem[W\" unsch et al. (2007)]{RWetal1}
    {W\" unsch, R., Silich, S., Palou\v s, J., \& Tenorio-Tagle, G.} 2007, 
    \textit{A\&A}, 471, 579

\bibitem[W\" unsch et al. (2008)]{RWetal2}
    {W\" unsch, R., Tenorio-Tagle, G., Palou\v s, J., \& Silich, S.} 2008, 
    \textit{ApJ}, 684, September 1, arXiv:0805.1380v1

\end{thebibliography}
\end{document}